\documentclass[11pt]{article}
\usepackage{epsfig,sint,cite,amsfonts}

\usepackage[english]{babel}
\usepackage{simplewick}

\newcommand{\be}{\begin{equation}}
\newcommand{\ee}{\end{equation}}
\newcommand{\bea}{\begin{eqnarray}}
\newcommand{\eea}{\end{eqnarray}}
\newcommand{\id}{1\!\!1}
\newcommand{\Pg}{\hat{{\rm P}}_{\rm G}}
\newcommand{\Tr}{\,\hbox{\rm Tr}}

\begin{document}

\begin{titlepage}
\begin{flushright}
\hfill CERN-PH-TH/2008-112\\
\end{flushright}

\vskip 2.5 cm
\begin{center}
  {\Large\bf Symmetries and exponential error reduction\\ 
             in Yang-Mills theories on the lattice\\[0.5ex]} 
\end{center}
\vskip 0.5 cm
\begin{center}
{\large Michele Della Morte and Leonardo Giusti}
\vskip 0.75cm
CERN, Physics Department, TH Division, CH-1211 Geneva 23, Switzerland
\vskip 3.0cm
{\bf Abstract}
\vskip 0.35ex
\end{center}

\noindent
The partition function of a quantum field theory with an exact symmetry can be 
decomposed into a sum of functional integrals each giving the contribution 
from states with definite symmetry properties. The composition rules
of the corresponding transfer matrix elements can be exploited to devise
a multi-level Monte Carlo integration scheme for 
computing correlation functions whose numerical cost, 
at a fixed precision and at asymptotically 
large times, increases power-like with the time extent of the lattice. 
As a result the numerical effort is exponentially reduced with respect 
to the standard Monte Carlo procedure. We test this strategy in the SU(3) 
Yang--Mills theory by evaluating the relative contribution to the partition 
function of the parity odd states.

\vfill

\eject

\vfill

\eject

\end{titlepage}

\section{Introduction}
Dynamical properties of quantum field theories can be 
determined on the lattice by computing appropriate functional 
integrals via Monte Carlo simulations. For the most interesting 
theories this is, up to now, the only tool to carry out non-perturbative 
computations from first principles. The mass of the lightest asymptotic 
state with a given set of quantum numbers can, for instance, be extracted from 
the Euclidean time dependence of a suitable two-point correlation 
function. Its contribution can be disentangled from those of other 
states by inserting the source fields at large-enough time distances.
The associated statistical error can be estimated 
from the spectral properties of the theory~\cite{Parisi:1983ae,Lepage:1989hd}. 
Very often the latter grows exponentially with the time separation, 
and in practice it is not possible to find a window where statistical and systematic 
errors are both under control. 
This is a well known major limiting factor in many numerical 
computations such as, for example, the computation of the glueball masses 
in the Yang--Mills theory.
A widely used strategy to mitigate this problem is to reduce the systematic 
error by constructing interpolating operators with a small overlap on the 
excited states~\cite{Albanese:1987ds,Teper:1987wt}. The lowest energy is 
then extracted at short time-distances by assuming a negligible contamination 
from excited states, sometimes also with the help of anisotropic
lattices~\cite{Morningstar:1997ff,Morningstar:1999rf}. This procedure is not entirely 
satisfactory from a conceptual and a practical point of view. The exponential problem 
remains unsolved, and the functional form of the sources are usually optimized so that 
the correlator shows a single exponential decay in the short time range allowed by 
the statistical noise. A solid evidence that a single state dominates the correlation 
function, i.e.  a long exponential decay over many orders of magnitude, is thus missing.

\indent In this paper we propose a computational strategy to solve the 
exponential problem. The latter arises in the standard procedure since
for any given gauge configuration all asymptotic states of the theory are 
allowed to propagate in the time direction, regardless of the quantum numbers of the 
source fields. By using the transfer matrix formalism, we introduce projectors 
in the path integral which, configuration by configuration, 
permit the propagation in time of states with a given set of  quantum numbers only. 
The composition properties of the projectors can then be exploited to implement a 
hierarchical multi-level integration procedure similar to those proposed 
in Refs.~\cite{Parisi:1983hm,Luscher:2001up} for the Polyakov loops. 
By iterating over several levels the numerical cost of computing the relevant 
observables grows, at asymptotically large times, with a power of the time 
extent of the lattice. 

\indent We test our strategy of a ``symmetry constrained'' Monte Carlo 
in the SU(3) Yang--Mills theory by determining the relative contribution to the 
partition function of the parity-odd states on lattices with a spacing of roughly 
$0.17$~fm, spatial volumes up to $2.5\, \mbox{fm}^3$, and time extent up 
to $3.4\, \mbox{fm}$. The algorithm behaves as expected, and in particular the 
multi-level integration scheme achieves an exponential reduction of the 
numerical effort. In the specific numerical implementation adopted here the computation 
of the projectors is the most expensive part, and its cost scales roughly with the square of the 
three-dimensional volume. The realistic lattices considered in this paper, 
however, were simulated with a modest computational effort. 

\indent The strategy proposed here is rather general and we 
expect it to be applicable to other symmetries and other field theories including 
those having fermions as fundamental degrees of freedom. It can, of course, be 
quite useful also for computing excited levels in other quantum 
mechanical systems. The basic ideas were indeed checked
in a considerable simpler and solvable quantum system with a non-trivial parity symmetry, 
namely the one dimensional harmonic oscillator~\cite{DellaMorte:2007zz}.

\section{Preliminaries and basic notation}
We set up the SU(3) Yang--Mills theory on a finite four-dimensional 
lattice of volume $V=T\times L^3$ with a spacing $a$ 
and periodic boundary conditions\footnote{Throughout the paper dimensionful quantities 
are always expressed in units of $a$.}. The gluons are discretized
through the standard Wilson plaquette action 
\be
S[U] = \frac{\beta}{2}\, \sum_{x} \sum_{\mu,\nu} 
\left[1 - \frac{1}{3}{\rm Re}\Tr\Big\{U_{\mu\nu}(x)\Big\}\right]\; ,
\ee
where the trace is over the color index, $\beta=6/g_0^2$ with $g_0$ 
the bare coupling constant, and the plaquette is defined as a function 
of the gauge links $U_\mu(x)$ as
\be\label{eq:placst}
U_{\mu\nu}(x) = U_\mu(x)\, U_\nu(x+\hat \mu)\, U^\dagger_\mu(x + \hat \nu)\,
             U^\dagger_\nu(x)\; ,  
\ee
with $\mu,\,\nu=0,\dots,3$, $\hat \mu$ is the unit vector along the 
direction $\mu$ and $x$ is the space-time coordinate. 
The action is invariant under a gauge transformation
\be\label{eq:GT}
U_\mu(x) \longrightarrow U^\Omega_\mu(x) = \Omega(x)\, U_\mu(x)\, 
\Omega^\dagger(x+\hat\mu)\; 
\ee
with $\Omega(x)\in{\rm SU(3)}$. The path integral is defined as usual
\be\label{eq:Zstand}
Z = \int {\rm D}_4 [U] \; e^{-S[U]}\; ,\qquad 
{\rm D}_4 [U] = \prod_{x}\, \prod_{\mu=0}^{3} {\cal D} U_\mu(x)\; ,
\ee
where ${\cal D} U$ is the invariant Haar measure on the SU(3) group, which throughout the paper 
will be always normalized such that $\int {\cal D} U = 1$. The average value of a generic 
operator ${\cal O}$ can thus be written as 
\be
\displaystyle
\langle {\cal O} \rangle = \frac{1}{Z} \int {\rm D}_4 [U]\; e^{-S[U]}\, {\cal O}[U]\; .
\ee

\subsection{Hilbert space}\label{sec:HS}
The Hilbert space of the theory is the space of all square-integrable 
complex-valued functions $\psi[V]$ of $V_k({\bf x})\in {\rm SU(3)}$
with a scalar product defined as (${\bf x}$ is the three dimensional 
space-coordinate and $k=1,2,3$) 
\be\displaystyle
\langle \phi | \psi\rangle = \int {\rm \bf D}_3[V] \, \phi[V]^* \psi[V] \; , \qquad
{\rm \bf D}_3[V] = \prod_{{\bf x}}\, \prod_{k=1}^{3} 
{\cal D} V_k({\bf x})\; .
\ee
The ``coordinate'' basis is the set of vectors which 
diagonalize the field operator at all points ${\bf x}$, i.e. 
\be
\hat {\rm V}_k({\bf x}) |V\rangle = V_k({\bf x}) |V\rangle\; ,
\ee
and which are normalized such that 
\be
\langle V | \psi \rangle = \psi[V] \; .
\ee
From a quantum mechanical point of view, the field 
values $V_k({\bf x})$ form the set of quantum numbers 
that label the vectors of the basis. In a gauge theory 
physical states are wave functions which satisfy
\be
\psi[V^\Omega] = \psi[V]
\ee
for all gauge transformations $\Omega$. A projector onto 
this subspace can be defined as 
\be\label{eq:Pg}
\langle V | \Pg | \psi \rangle = \int {\rm \bf D}[\Omega] \, \psi[V^\Omega]\; ,
\qquad
{\rm \bf D}[\Omega] = 
\prod_{{\bf x}} {\cal D}\Omega({\bf x})\; , 
\ee
and it is straightforward to verify that $\Pg^2=\Pg$. 

\subsection{Transfer matrix}
The transfer matrix of a Yang-Mills theory discretized 
by the Wilson action has been constructed many years 
ago~\cite{Wilson:1977nj,Luscher:1976ms,Creutz:1976ch,Osterwalder:1977pc}.  
The subject is well known and it appears on text books, therefore
we report only those formul\ae \  which are relevant to 
the paper. The starting point is to rewrite the functional integral in 
Eq.~(\ref{eq:Zstand}) as 
\be\label{eq:Z1}
Z = \int \prod_{x_0=0}^{T-1}\, 
{\rm \bf D}_3[V_{x_0}]\, {\rm T}\Big[V_{x_0+1},V_{x_0}\Big]
\ee
where the transfer matrix elements are defined as 
\be
{\rm T}\Big[V_{x_0+1},V_{x_0}\Big] = \int {\rm \bf D}[\Omega]
\; e^{-L[V^{\Omega}_{x_0+1},V_{x_0}]} \; ,
\ee
with 
\be
L\Big[V_{x_0+1},V_{x_0}\Big] =  K\Big[V_{x_0+1},V_{x_0}\Big]
+  \frac{1}{2}W\Big[V_{x_0+1}\Big] + \frac{1}{2}W\Big[V_{x_0}\Big]\; ,
\ee
and $\Omega$ being identified with the link in the temporal direction. 
The kinetic and the potential contributions to the Lagrangian are 
given by
\be
K\Big[V_{x_0+1},V_{x_0}\Big] = \beta \sum_{{\bf x},k} 
\left[ 1 - \frac{1}{3}{\rm Re}\Tr\left\{ 
V_{k}(x_0+1,{\bf x}) V^\dagger_{k}(x_0,{\bf x}) 
\right\}\right]\; ,
\ee
and 
\be
W\Big[V_{x_0}\Big] = \frac{\beta}{2} \sum_{{\bf x}} \sum_{k,l} 
\left[1 - \frac{1}{3}{\rm Re}\Tr\Big\{V_{kl}(x_0,{\bf x})\Big\}\right]\; , 
\ee
respectively, where $V_{kl}$ is the plaquette defined in Eq.~(\ref{eq:placst})
computed with the links $V_k({\bf x})$. The potential term is gauge-invariant, i.e.
$W\Big[V_{x_0}\Big] = W\Big[V^{\Omega}_{x_0}\Big]$, while the dependence of the 
kinetic term on the gauge transformations 
$\Omega'$ at time $(x_0+1)$ and $\Omega$ at time $x_0$
is only via the product $\Omega^\dagger\Omega'$. Thanks to 
the invariance of the Haar measure under left and right multiplication,
this implies that the transfer matrix is gauge-invariant 
\be
{\rm T}\Big[V_{x_0+1}^{\Omega'},V^{\Omega}_{x_0}\Big] = 
{\rm T}\Big[V_{x_0+1},V_{x_0}\Big]\; ,
\ee
and that
\be\label{eq:Tgg}
{\rm T}\Big[V_{x_0+1},V_{x_0}\Big] = \int {\rm \bf D}[\Omega'] 
{\rm \bf D}[\Omega] 
\; e^{-L[V^{\Omega'}_{x_0+1},V^{\Omega^\dagger}_{x_0}]} \; .
\ee
The latter are thus  
matrix elements of a transfer operator $\hat {\rm T}$ 
between gauge invariant states
\be
{\rm T}\Big[V_{x_0+1},V_{x_0}\Big] = 
\left\langle V_{x_0+1}| \Pg \hat {\rm T}\Pg | V_{x_0}\right\rangle\; , 
\ee
and the functional integral can then be written as  
\be\label{eq:Z2}
Z = \Tr \left\{ \left[\hat {\rm T}\Pg\right]^T \right\}\; ,
\ee 
where the trace is over all gauge invariant states. For a thick time-slice, 
i.e. the ensemble of points in the sub-lattice with time
coordinates in a given interval $[x_0,y_0]$ and bounded by the equal-time
hyper-planes at times $x_0$ and $y_0$, the transfer matrix elements can be
introduced by the formula
\be\label{eq:T1}
{\rm T}\Big[V_{y_0},V_{x_0}\Big] = \int \prod_{w_0=x_0+1}^{y_0-1}\, 
{\rm \bf D}_3[V_{w_0}]\, 
\prod_{z_0=x_0}^{y_0-1} {\rm T}\Big[V_{z_0+1},V_{z_0}\Big]\; .
\ee

\section{Decomposition of the functional integral}
The invariance of the system under a global symmetry 
can be exploited to decompose the partition function
into a sum of functional integrals each giving the contribution 
from states with definite symmetry properties. In the following we will 
focus on the invariance of the Yang--Mills theory under parity.\\ 
\indent In the coordinate basis, the parity transformation on gauge 
invariant states can be defined as 
\be\label{eq:PT}
\mbox{\LARGE $\hat\wp$}\,|{\rm V}\rangle = |{\rm V}^\wp\rangle\; , \qquad  
|{\rm V}\rangle = \Pg |V\rangle\; ,  \qquad V^\wp_k({\bf x}) = V^\dagger_k(-{\bf x}-\hat k)\; ,
\ee
which implies that {\Large $\hat {\wp}^2$}=$\id$.
The parity eigenstates can then be written as 
\be
|{\rm V},\pm\rangle = \frac{1}{\sqrt{2}}\Big[|{\rm V}\rangle \pm |{\rm V}^\wp\rangle \Big]\;, \qquad  
\mbox{\LARGE $\hat\wp$}\, |{\rm V},\pm\rangle = \pm |{\rm V},\pm\rangle\; .
\ee
and their transfer matrix elements are given by
\bea
\langle s', {\rm V}_{x_0+1}| \hat {\rm T} | {\rm V}_{x_0}, s\rangle & = & 2\,
\delta_{s's}\; {\rm T}^s\Big[V_{x_0+1},V_{x_0}\Big]\; ,\\[0.25cm]
{\rm T}^s\Big[V_{x_0+1},V_{x_0}\Big] & = & \frac{1}{2}
\left\{ {\rm T}\Big[V_{x_0+1},V_{x_0}\Big] + s\, 
        {\rm T}\Big[V_{x_0+1},V^\wp_{x_0}\Big] \right\}\; .\label{eq:Ts}
\eea
The invariance of the action yields  
\be\label{eq:Tp}
{\rm T}\Big[V^\wp_{x_0+1},V^\wp_{x_0}\Big] = 
{\rm T}\Big[V_{x_0+1},V_{x_0}\Big]\;, 
\qquad {\rm T}\Big[V^\wp_{x_0+1},V_{x_0}\Big] = 
{\rm T}\Big[V_{x_0+1},V^\wp_{x_0}\Big]\; ,
\ee
and therefore
\be\label{eq:Tsp}
{\rm T}^s\Big[V_{x_0+1},V^\wp_{x_0}\Big] = s\, {\rm T}^s\Big[V_{x_0+1},V_{x_0}\Big]\; .
\ee
For a thick time-slice the matrix elements between parity states
can be introduced by exploiting the composition rule
\be
{\rm T}^s\Big[V_{y_0},V_{x_0}\Big] = 
\left\{{\rm T}^s\Big[V_{y_0},V_{z_0}\Big]\; 
       {\rm T}^s\Big[V_{z_0},V_{x_0}\Big]\right\}
\; ,\label{eq:comp1} 
\ee
where $x_0 < z_0 < y_0$ and in general
\be
\left\{{\rm T}^s\Big[V_{y_0},V_{z_0}\Big]\;
       {\rm T}^{s'}\Big[V_{z_0},V_{x_0}\Big]
\right\} = \int {\rm \bf D}_3[V_{z_0}]\,
{\rm T}^s\Big[V_{y_0},V_{z_0}\Big]\,  
{\rm T}^{s'}\Big[V_{z_0},V_{x_0}\Big]\; .
\ee
It is easy to show that, 
in addition to relations analogous to those in 
Eqs.~(\ref{eq:Ts})--(\ref{eq:Tsp}), the identities
\bea
\left\{{\rm T}^s\Big[V_{y_0},V_{z_0}\Big]\; 
{\rm T}^{-s}\Big[V_{z_0},V_{x_0}\Big]\right\} & = & 0\; ,\label{eq:comp2}\\[0.25cm]
\left\{{\rm T}^s\Big[V_{y_0},V_{z_0}\Big]\; 
       {\rm T}\Big[V_{z_0},V_{x_0}\Big]\right\} & = & 
{\rm T}^s\Big[V_{y_0},V_{x_0}\Big]\label{eq:comp3}
\eea
hold. In particular they imply that  
\be\label{eq:Z3}
 \frac{{\rm T}^s[V_{y_0},V_{x_0}]}
      {{\rm T}\;[V_{y_0},V_{x_0}]}= 
\frac{1}{Z_\mathrm{sub}}\int 
 {\rm D}_4 [U]_\mathrm{sub} \; e^{-S[U]}\, 
\frac{{\rm T}^s[U_{y_0},U_{y_0-1}]}
     {{\rm T}\,[U_{y_0},U_{y_0-1}]}\; ,
\ee
an useful expression for the practical implementation of the 
multi-level algorithm described in the following section.
The subscript ``sub'' indicates that the integral is performed over the 
dynamical field variables in the thick time-slice $[x_0,y_0]$ with 
the spatial components $U_k(x)$ of the boundary fields 
fixed to $V_k(x_0,\vec {\bf x})$ and $V_k(y_0,\vec {\bf x})$ respectively.
Finally, by inserting Eq.~(\ref{eq:Ts}) into Eq.~(\ref{eq:Z1})
and repeatedly applying Eq.~(\ref{eq:comp2}), it is possible to rewrite the 
path integral as a sum of functional integrals 
\be
Z = \sum_{s=\pm} Z^s\; , \qquad Z^s = \int \prod_{x_0=0}^{T-1}\, 
{\rm \bf D}_3[V_{x_0}]\, {\rm T}^s\Big[V_{x_0+1},V_{x_0}\Big]\; ,
\ee
each giving the contribution from gauge-invariant parity-even 
and -odd states respectively
\be\label{eq:exp2}
Z^+ = e^{-E_0\, T}  \left[1 + \sum_{n=1} w^+_n\, e^{-E_n^+ T}\right]\, , 
\qquad  Z^- = e^{-E_0\, T} \sum_{m=1} w^-_m\, e^{-E_m^- T}\; .
\ee
In these expressions $E_0$ is the vacuum energy, $E_n^+$ and 
$E_m^-$ are the energies (with respect to the vacuum one)
of the parity even and odd eigenstates, and $w^+_n$ and $w^-_m$ are 
the corresponding weights. The latter are integers and positive since 
for the Wilson action the transfer operator $\hat {\rm T}$ is self-adjoint 
and strictly positive~\cite{Luscher:1976ms}.

It is interesting to notice that
even though the transfer matrix formalism inspired the construction, 
the above considerations hold independently of the existence of a 
positive self-adjoint transfer operator. The insertion of 
${\rm T}^s [V_{y_0},V_{x_0}]$ in the path integral plays the r\^ole of 
a projector, as on each configuration it allows the propagation in the 
time direction of states with parity $s$ only. Indeed the parity transformation 
of one of the boundary fields in ${\rm T}[V_{y_0},V_{x_0}]$ 
flips the sign of all contributions that it receives from the parity-odd 
states while leaving invariant the rest. The very same applies 
to the path integral in Eq.~(\ref{eq:Zstand}) if the periodic boundary
conditions are replaced by $\wp$-periodic boundary conditions, i.e.
$V_T=V^\wp_0$. All contributions from the parity odd states are 
then multiplied by a minus sign. Similar 
considerations have already been exploited in different contexts, 
for instance in the study of the interface free energy of
the three-dimensional Ising model~\cite{Caselle:2007yc}.

\section{Multi-level simulation algorithm}
The composition rules in Eqs.~(\ref{eq:comp1})--(\ref{eq:comp3}) 
are at the basis of our strategy for computing $Z^s/Z$ (as well
as a generic correlation function) with a hierarchical multi-level 
integration procedure.

\subsection{Projector computation}
To determine the parity projector between two 
boundary fields of a thick time-slice, the basic building block 
to be computed is the ratio of transfer matrix elements 
\be\label{eq:rat}
{\rm R}[V_{x_0+d},V_{x_0}] =  
\frac{{\rm T}[V_{x_0+d},V^\wp_{x_0}]}{{\rm T}[V_{x_0+d},V_{x_0}]}\; .
\ee
The parity transformation in the numerator changes one of the boundary fields 
over the entire spatial volume of the corresponding time-slice, 
a global operation which could make the logarithm of this ratio 
proportional to the spatial volume, see for instance~\cite{Caselle:2007yc}. 
The transfer matrix formalism and the expected spectral properties of the 
Yang--Mills theory however suggest that, in a finite volume and for $d$ large enough,  
only a few of the physical states give a sizeable contribution to this ratio, 
which is therefore expected to be of $O(1)$. These general properties
can be studied analytically for the free lattice scalar theory, see for 
instance~\cite{Rothe:1997kp}. It goes without saying that the latter has 
a different spectrum from the Yang--Mills theory, and therefore can be used only 
as an example where our strategy can be studied analytically.\\
\indent Even tough the ratio $R$ is expected to be of $O(1)$, 
the integrands in the numerator and in the denominator on the r.h.s of 
Eq.~(\ref{eq:rat}) are, in general, very different and the main
contributions to their integrals come from different regions of the phase space. 
The most straightforward way for computing $R$ is to define a set of $n$ 
systems with partition functions ${\cal Z}_1\, \dots\, {\cal Z}_n$ 
designed in such a way that the relevant phase spaces of successive integrals 
overlap and that ${\cal Z}_1={\rm T}[V_{x_0+d},V^\wp_{x_0}]$ and 
${\cal Z}_n={\rm T}[V_{x_0+d},V_{x_0}]$. The ratio $R$ can then be calculated
as 
\be
{\rm R} = 
\frac{{\cal Z}_1}{{\cal Z}_2} \times \frac{{\cal Z}_2}{{\cal Z}_3}
\times \dots
\times \frac{{\cal Z}_{n-2}}{{\cal Z}_{n-1}} 
\times \frac{{\cal Z}_{n-1}}{{\cal Z}_n} \; ,
\ee
with each ratio on the r.h.s. being computable in a single 
Monte Carlo simulation by averaging the proper reweighting factor. 
To implement this procedure we start by generalizing the definition of the 
transfer matrix element in 
Eq.~(\ref{eq:Tgg}) as 
\be\label{eq:Tggr}
\overline {\rm T}\Big[V_{x_0+1},V_{x_0},r\Big] = \int {\rm \bf D}[\Omega'] 
{\rm \bf D}[\Omega] 
\; e^{-\overline L[V^{\Omega'}_{x_0+1},V^{\Omega^\dagger}_{x_0},r]} \; ,
\ee
where $r\in[-1/2,1/2]$ and 
\bea\label{eq:kin2}\displaystyle
\overline L\Big[V_{x_0+1},V_{x_0},r\Big] & = &  \Big(\,\frac{1}{2}+r\Big)\, 
K\Big[V_{x_0+1},V_{x_0}\Big]
+ \Big(\,\frac{1}{2} - r\Big)\, K\Big[V^\wp_{x_0+1},V_{x_0}\Big]\nonumber\\[0.25cm]
& + & \frac{1}{2}W\Big[V_{x_0+1}\Big] + \frac{1}{2}W\Big[V_{x_0}\Big]\; .
\eea
Analogously, Eq.~(\ref{eq:Z3}) can be generalized as 
\be\label{eq:Zdtw}
\overline {\rm T}\Big[V_{x_0+d},V_{x_0},r\Big] = \int \prod_{w_0=x_0+1}^{x_0+d-1}\, 
{\rm \bf D}_3[V_{w_0}]\, 
\left[\prod_{z_0=x_0}^{x_0+d-2}\, {\rm T}\Big[V_{z_0+1},V_{z_0}\Big]\right]\, 
\overline {\rm T}\Big[V_{x_0+d},V_{x_0+d-1},r\Big]
\ee
and the ratio ${\rm R}[V_{x_0+d},V_{x_0}]$ can be written as 
\be\displaystyle\label{eq:bigP}
{\rm R}[V_{x_0+d},V_{x_0}] = 
\prod_{k=1}^{L^3}
{\overline{\rm R}}[V_{x_0+d},V_{x_0},-1/2 + (k-1/2)\,\varepsilon]
\ee
where 
\be\label{eq:rat2}
{\overline{\rm R}}[V_{x_0+d},V_{x_0},r] = 
\frac{\overline{\rm T}[V_{x_0+d},V_{x_0},r-\,\varepsilon/2]}
     {\overline{\rm T}[V_{x_0+d},V_{x_0},r+\,\varepsilon/2]}
\ee
and $\varepsilon=1/L^3$. With this choice of $\varepsilon$ 
the relevant phase spaces of two consecutive integrals overlap
since the actions differ by a quantity of $O(1)$, while their 
fluctuations are of $O(\sqrt{V})$. To compute each ratio on the r.h.s. of 
Eq.~(\ref{eq:bigP}) one starts by noticing that 
the group integrals on $\Omega^{'}$ and $\Omega$
in Eq.~(\ref{eq:Tggr}) can be factorized by introducing on each point of the 
time-slice $x_0$ the usual temporal link 
$U_0(x_0,\vec x)= \Omega^\dagger(\vec x) \Omega^{'}(\vec x)$ and a second 
temporal link $U_4(x_0,\vec x)=\Omega^\dagger(\vec x) \Omega^{'}(-\vec x)$. 
The average of the reweighting factor is then computed 
with the three-level algorithm described in Appendix \ref{app:appA}.
As other methods for computing ratios of partition 
functions which are present in the 
literature~\cite{Ferrenberg:1989ui,Hoelbling:2000su,deForcrand:2000fi}, the numerical cost 
scales roughly quadratically with the three-dimensional volume. Since the main goal 
of this paper is to present and test the validity of the strategy, we leave to future 
studies the development of a more refined 
and better scaling algorithm for the computation of the projector.

\subsection{Hierarchical integration}\label{sec:HI}
Once the projectors have been computed, the ratio of partition functions
$Z^s/Z$ can be calculated by implementing the hierarchical two-level 
integration formula
\be\label{eq:bella}
\frac{Z^s}{Z} = \frac{1}{Z}\int {\rm D}_4 [U]\, e^{-S[U]} 
\, {{\rm P}}^s_{m,d}\Big[T,0\Big]
\ee
where 
${\rm P}^s_{m,d}\Big[y_0,x_0\Big]$ is defined as 
\be
\displaystyle\label{eq:Psmd}
{{\rm P}}^s_{m,d}\Big[y_0,x_0\Big] =  
{\prod_{i=0}^{m-1}} 
\frac{{\rm T^s}[U_{x_0+(i+1)\cdot d},U_{x_0 + i\cdot d}]}
     {{\rm T}[U_{x_0+(i+1)\cdot d},U_{x_0 + i\cdot d}]}
\ee
with $m\ge 1$ and $y_0=x_0 + m \cdot d$. The procedure can, of course, 
be generalized to a multi-level
algorithm. For a three-level one, for instance, each ratio on the 
r.h.s of Eq.~(\ref{eq:Psmd}) can be computed by a 
two-level scheme. Thanks to the 
composition rules in Eqs.~(\ref{eq:comp1}) and (\ref{eq:comp2}),
the r.h.s. of Eq.~(\ref{eq:bella}) does not depend on 
$m$ and $d$. When computed by a Monte Carlo procedure, however,
its statistical error depends strongly on the specific
form of ${{\rm P}}^s_{m,d}\Big[y_0,x_0\Big]$ chosen. 
The algorithm therefore requires an optimization 
which in general depends on the spectral properties of the 
theory. It is however important to stress that the 
multi-level hierarchical integration gives always the correct 
result independently on the details of its implementation. This 
can be shown by following the same steps in the Appendix A of 
Ref.~\cite{Luscher:2001up}. There are two main differences: 
auxiliary link variables and their own actions need to be introduced for each value
of $r$, and the computation of ${\overline{\rm R}}$ 
requires a thermalization procedure for each value of $r$. We do not 
expect the latter to be particularly problematic since, as mentioned earlier, 
expectation values for consecutive values of $r$ refer to path integrals 
with the relevant phase spaces which overlap. The ratios ${\overline{\rm R}}$ 
are computed by simulating 
systems corresponding to consecutive values of $r$ one after the other, and by 
starting from the one used to extract the boundary fields ($r=0.5$).

\subsection{Exponential error reduction}
The statistical variance of the estimate of a two-point 
correlation function $\langle  O(x_0) O(0) \rangle$
of a parity-odd interpolating 
operator $O$, computed by the standard Monte Carlo procedure, is defined as
\be
\sigma^2 = \langle O^2(x_0)  O^2(0) \rangle  - \langle O(x_0) O(0) \rangle^2 \; .
\ee
At asymptotically-large time separations the signal-to-noise 
ratio can be easily computed via the transfer matrix formalism which, 
for $0\ll x_0 \ll T/2$, gives~\cite{Parisi:1983ae,Lepage:1989hd}
\be\displaystyle
\frac{\langle  O(x_0) O(0) \rangle}
     {\sigma} = 
\frac{|\langle E_1^-|\hat  O | 0 \rangle|^2}
     {|\langle 0 |\hat O^2 | 0 \rangle|}\,
     e^{- E_1^- x_0} + \cdots
\ee 
The exponential decrease of this ratio with the time distance 
can be traced back to the fact that for each gauge configuration 
the standard Monte Carlo allows for the propagation in 
time of all asymptotic states of the theory regardless 
of the quantum numbers of the source field $O$. Therefore each 
configuration gives a contribution to the signal which decreases 
exponentially in time, whereas it contributes $O(1)$ to the noise (variance) 
at any time distance. On the contrary, if in Eq.~(\ref{eq:bella})
$d$ is chosen large enough for the single thick-slice ratio to be 
roughly dominated by the contribution of the lightest state, then each factor 
is of order $e^{-E^-_1\, d}$. For each configuration of the boundary fields, 
the magnitude of the product is proportional to $e^{-E^-_1\, T}$, and the statistical 
fluctuations are reduced to this level. To achieve an analogous exponential gain 
in the computation of the correlation functions, the projectors ${\rm T^s}$ have to
be inserted in the proper way among the interpolating operators (see 
Ref.~\cite{DellaMorte:2007zz} for a more detailed discussion).

\begin{table}[!t]
\begin{center}
\begin{tabular}{|cccccc|}
\hline
Lattice&$L$&$T$&$N_\mathrm{conf}$&$N_\mathrm{lev}$&$d$\\[0.125cm]
\hline
${\rm A}_1$&   6  &  4  & 50 & 2 & 4 \\[0.125cm]
${\rm A}_2$&      &  5  & 50 & 2 & 5 \\[0.125cm]
${\rm A}_3$&      &  6  & 50 & 2 & 6 \\[0.125cm]
${\rm A}_4$&      &  8  & 175& 2 & 4 \\[0.125cm]
${\rm A}_5$&      & 10  & 50 & 2 & 5 \\[0.125cm]
${\rm A}_6$&      & 12  & 90 & 2 & 6 \\[0.125cm]
${\rm A}_7$&      & 16  & 48 & 2 & 8 \\[0.125cm]
${\rm A}_8$&      & 20  & 48 & 3 & \{5,10\}\\[0.125cm]
\hline
${\rm B}_1$&   8  &  4  & 20 & 2 & 4 \\[0.125cm]
${\rm B}_2$&      &  5  & 25 & 2 & 5 \\[0.125cm]
${\rm B}_3$&      &  6  & 75 & 2 & 3 \\[0.125cm]
${\rm B}_4$&      &  8  & 48 & 2 & 4 \\[0.125cm]
\hline
\end{tabular}
\caption{Simulation parameters: $N_\mathrm{conf}$ is the number of configurations
of the uppermost level, $N_\mathrm{lev}$ is the number of levels and $d$ is the thickness 
of the thick time-slice used for the various levels.\label{tab:lattices}}
\end{center}
\end{table}

\section{Numerical simulations}
We have tested the hierarchical multi-level integration strategy 
described in the previous section for the SU(3) Yang--Mills theory 
by performing extensive numerical computations. We have simulated lattices 
with an inverse gauge coupling of $\beta=6/g^2_0=5.7$ which corresponds  
to a value of the reference scale $r_0$ of about $2.93a$ 
\cite{Guagnelli:1998ud,Necco:2001xg}. The number of lattice points in each 
spatial direction has been set to $L=6,8$ corresponding to a linear 
size of $1.0$ and $1.4$~fm respectively. For each spatial volume we have 
considered several time extents $T$, the full list is 
reported in Table~\ref{tab:lattices} together with the number of 
configurations generated and the details of the multi-level 
simulation algorithm used for each run. The lattices have been 
chosen to test the strategy in a realistic situation with the computational 
resources at our disposal, i.e. a machine equivalent to approximatively 
6 dual processor quad-core PC nodes of the last generation running for a few
months.
\begin{figure}[!t]
\begin{center}
\includegraphics[width=12.0cm]{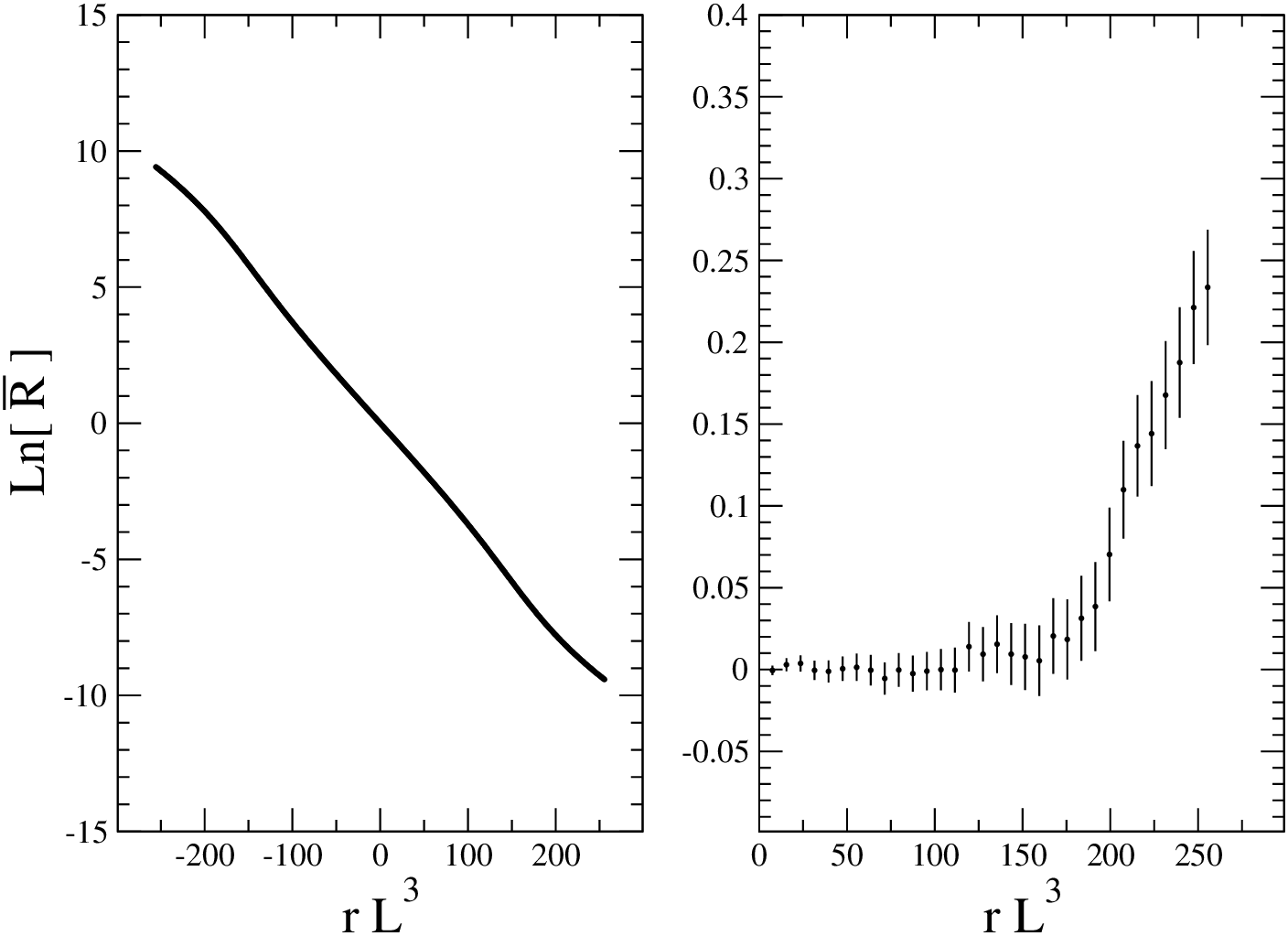}
\caption{Left: the natural logarithm of 
${\overline{\rm R}}[V_{x_0+d},V_{x_0},r]$ is shown as a function of 
$r$ (statistical errors are smaller than symbols) for a typical configuration 
of the run ${\rm B}_3$. 
Right: the sum of the points in the interval $[-r,r]$ is plotted 
as a function of $r$ (one each eighth point for visual 
convenience).\label{fig:freeE}}
\end{center}
\end{figure}

\subsection{Algorithm implementation and tests\label{sec:impl}}
The basic Monte Carlo update of each link variable is a combination of 
heatbath and over-relaxation updates which implements the 
Cabibbo--Marinari scheme~\cite{Cabibbo:1982zn}. Depending on the value of 
the coupling constant associated to the link at a given stage 
of the simulation, the heatbath updates the SU(2) sub-matrices 
by the Metropolis, the Creutz~\cite{Creutz:1980zw} or the 
Fabricius--Haan~\cite{Fabricius:1984wp,Kennedy:1985nu} algorithm. 
In the uppermost level the generation of the gauge field configurations consumes
a negligible amount of computer time. At this level we perform many update
cycles between subsequent configurations (typically 500 iterations of 1 heatbath and 
$L/2$ over-relaxation updates of all link variables) so that they can be assumed 
to be statistically independent. On each of these configurations we compute 
the ``observables'' ${{\rm P}}^s_{m,d}[T,0]$, with the most expensive part being 
the estimate of the thick-slice ratio ${\rm R}[V_{x_0+d},V_{x_0}]$ at the 
lowest algorithmic level. The latter is computed by using the 
three-level algorithm described in the previous section, with 
the parameter values tuned sequentially level by level 
so to minimize the actual CPU cost for the required statistical precision. 
In all runs this has been set to be 
at most $30\%$ of the expected absolute value of the deviation of $R$ from 1, 
the latter being determined by some preliminary exploratory tests. As mentioned in 
section \ref{sec:HI}, the algorithm requires a thermalization step for each value 
of $r$ which has been fixed, after several exploratory runs, to 500 sweeps 
of the full sub-lattice.\\ 
\begin{figure}[!t]
\begin{center}
\includegraphics[width=12.0cm]{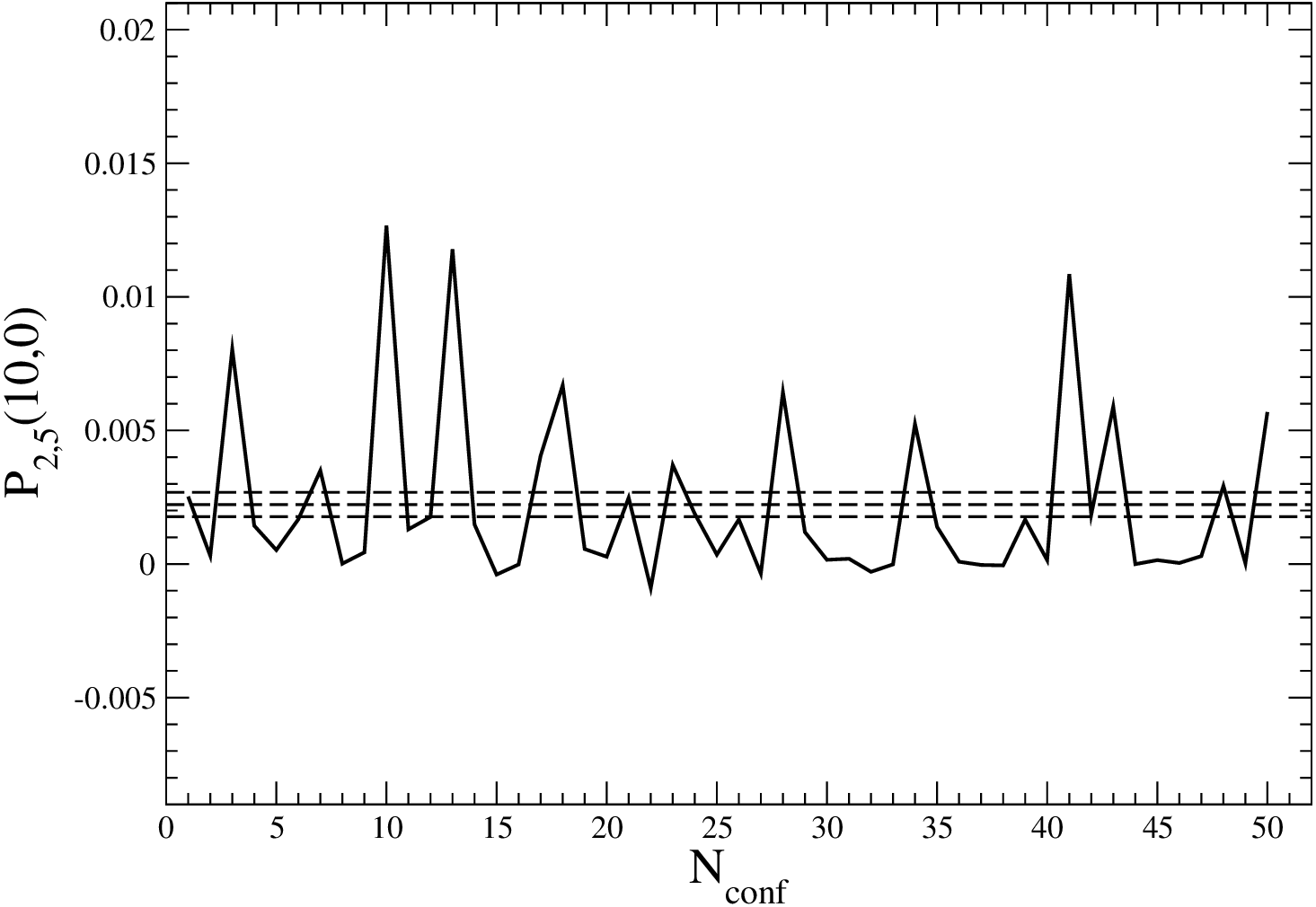}
\caption{Monte Carlo history of the quantity ${{\rm P}}^-_{2,5}[10,0]$ for the run 
${\rm A}_5$. The central dashed line corresponds to the average value, 
while the other two delimit the one standard deviation region.\label{fig:hist}}
\end{center}
\end{figure}
\indent Apart from many consistency checks of the programs, we have verified 
several non-trivial properties of the basic ratios 
in Eqs.~(\ref{eq:rat}) and (\ref{eq:rat2}). We have monitored 
the deviation from the equality
\be
{\overline{\rm R}}\Big[V_{x_0+d},V^\wp_{x_0},r\Big] = 
{\overline{\rm R}}\Big[V_{x_0+d},V_{x_0},-r\Big]^{-1}
\ee
for several boundary configurations and all values of $r$, and 
it turns out to be compatible with being a Gaussian statistical 
fluctuation. For the runs with $d=T$ we have verified that,
on each configuration and within the statistical error, 
the ratio ${\rm T^-}[V_T,V_0]/{\rm T}[V_T,V_0]$ 
is always positive as predicted by the 
transfer matrix representation. For $d=T/2$ the two thick-slice ratios 
in Eq.~(\ref{eq:bella}) have to be equal. We have monitored the difference 
in a significant sample of our configurations, and it turns out to be 
compatible with a Gaussian statistical fluctuation as well.\\
\indent The natural logarithm of  ${\overline{\rm R}}[V_{x_0+d},V_{x_0},r]$
is shown as a function of $r$ in the left panel of Fig.~\ref{fig:freeE}
for a typical configuration of the run ${\rm B}_3$. As expected, 
its value is of $O(1)$ for each value of $r$. Its almost
perfect asymmetry under $r\rightarrow -r$, however, makes the sum of all
the $L^3$ points a quantity of $O(1)$. This impressive cancellation,
which is at work for $T>3$ on both volumes, 
can be better appreciated in the right panel of the same Figure, where
the sum of the function in the interval $[-r,r]$ is plotted 
for a subset of values of $r$. It is the deviation from the exact 
asymmetry which flips in sign under a parity transformation of one of 
the boundary fields, and forms the signal we are interested in.
A similar behaviour 
is observed for all other configurations and runs.\\
\indent The Monte Carlo history of ${{\rm P}}^-_{2,T/2}[T,0]$ 
is shown in Figure~\ref{fig:hist} for the lattice ${\rm A}_5$. Also for 
all other runs we have observed
reasonable Monte Carlo histories, and therefore we have computed $Z^s/Z$
and its statistical error in the standard way. The run ${\rm A}_4$ 
however is much noisier than the others, with rather 
large fluctuations due to a few configurations. This could be 
related to the fact that $d=4$ is not yet large enough, and 
sizeable contaminations from the heavier states amplify the statistical 
fluctuations. To check our statistical errors, we have also carried out 
a more refined analysis following Ref.~\cite{Wolff:2003sm}. 
No autocorrelations among configurations have been observed, and the 
errors are fully compatible with those of the standard analysis.\\ 
\indent Before describing the 
main numerical results of the paper we mention that, for the runs
where $m=2$ is available, we have computed the quantity on the 
r.h.s of Eq.~(\ref{eq:comp2}). As expected, it turns out to be 
always compatible with zero.
\begin{table}[!t]
\begin{center}
\setlength{\tabcolsep}{.18pc}
\begin{tabular}{|c|cc|cc|cc|c|}
\hline
Lattice&$\displaystyle\frac{Z^+_{1,T}}{Z}$&$\displaystyle\frac{Z^-_{1,T}}{Z}$&
$\displaystyle\frac{Z^+_{1,T/2}}{Z}$&$\displaystyle\frac{Z^+_{2,T/2}}{Z}$&
$\displaystyle\frac{Z^-_{1,T/2}}{Z}$&$\;\;\;\; \displaystyle\frac{Z^-_{2,T/2}}{Z}$&
$M^-$\\[0.25cm]
\hline
${\rm A}_1$& 0.591(8) & 0.409(8)  & - & - & - & - & 0.223(5)  \\[0.125cm]
${\rm A}_2$& 0.823(13)& 0.177(13) & - & - & - & - & 0.346(14) \\[0.125cm]
${\rm A}_3$& 0.931(7) & 0.069(7)  & - & - & - & - & 0.446(17) \\[0.125cm]
${\rm A}_4$& -        & -         & 0.995(9) & 1.004(20) & 0.005(9) & $1.47 (28) \cdot 10^{-2}$ & 0.528(24) \\[0.125cm]
${\rm A}_5$& -        & -         & 1.003(7)  & 1.009(14) &-0.003(7)  & $2.2 \,\,(5)\,\, \cdot 10^{-3}$  & 0.611(20) \\[0.125cm]
${\rm A}_6$& -        & -         & 0.998(3)  & 0.996(5)  & 0.002(3)  & $6.6 \, (17)\, \cdot 10^{-4}$ & 0.610(21)\\[0.125cm]
${\rm A}_7$& -        & -         & 1.0006(9)& 1.0012(17) & -0.0006(9) & $2.8 \,\,(8)\,\, \cdot 10^{-5}$& 0.655(18) \\[0.125cm]
${\rm A}_8$& -        & -         & 0.9988(20) & 0.998(4) & 0.00024(20) & $1.5\,\,(5)\,\, \cdot 10^{-6}$& 0.670(15) \\[0.125cm]
\hline
${\rm B}_1$& 0.574(8) & 0.426(8)  & - & - & - & - & 0.213(5) \\[0.125cm]
${\rm B}_2$& 0.939(6) & 0.061(6)& - & - & - & - & 0.558(21)\\[0.125cm]
${\rm B}_3$& -        & -         & 0.979(15) & 0.97(3) & 0.021(15) & $1.65(26) \cdot 10^{-2}$ & 0.685(27)\\
${\rm B}_4$& -        & -         & 0.997(5) & 0.995(11) & 0.003(5) & $1.37 (26)\cdot 10^{-3}$ & 0.824(24)\\

\hline
\end{tabular}
\caption{Numerical results for various primary observables 
and for $M^-$ (see text).\label{tab:results}}
\end{center}
\end{table}

\subsection{Simulation results}
The ratios $Z^s/Z$ have been computed 
for all values of $m$ available in each run by using Eq.~(\ref{eq:bella}). 
The results are collected in Table~\ref{tab:results}, where they are identified by 
the obvious notation $Z^s_{m,d}/Z$.\\ 
\begin{figure}[!t]
\begin{center}
\includegraphics[width=12.0cm]{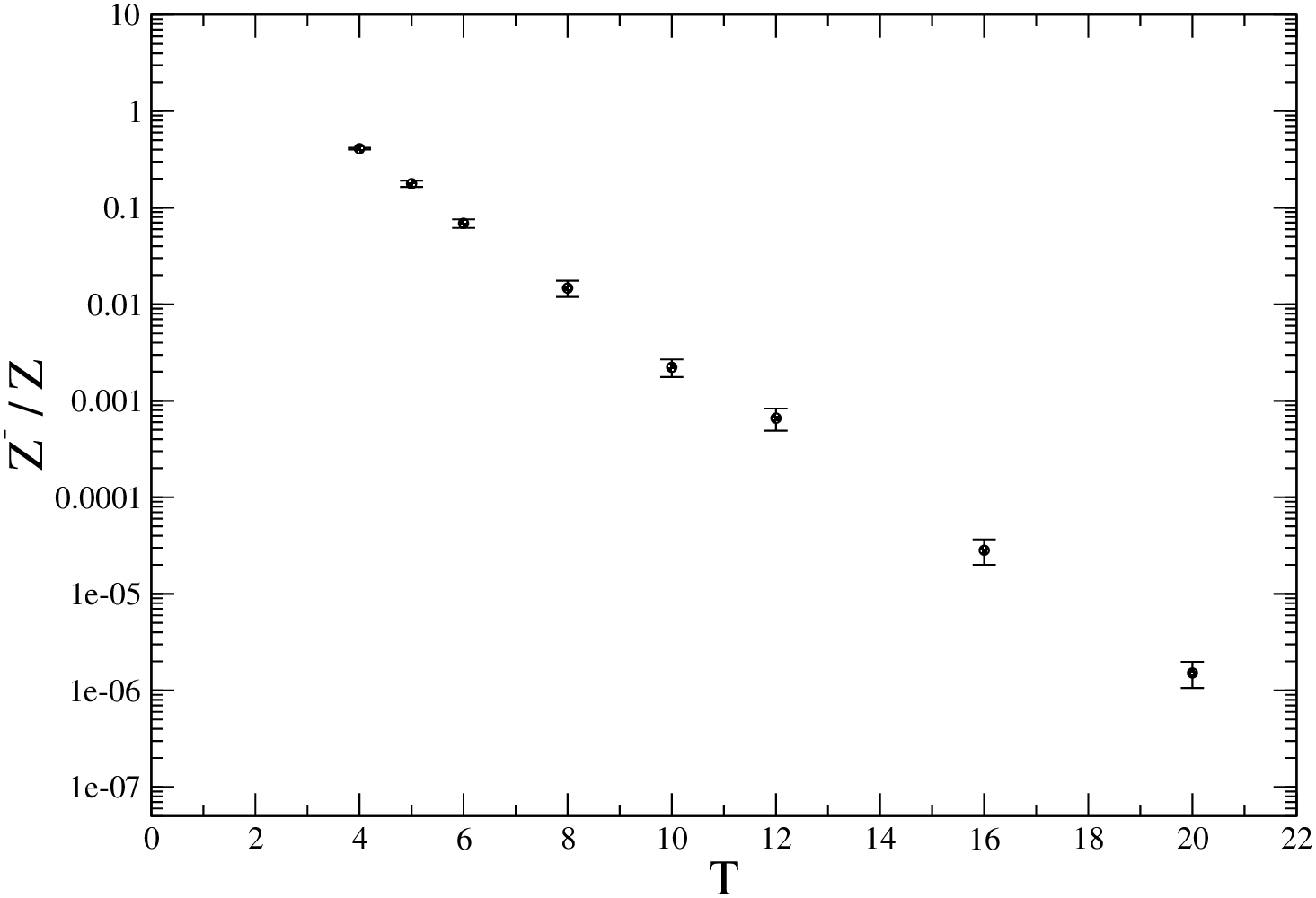}
\caption{The quantity $Z^-/Z$ as a function of $T$.\label{fig:ZmZp}}
\end{center}
\end{figure}
\begin{figure}[!t]
\begin{center}
\vspace{2.0cm}
\includegraphics[width=12.0cm]{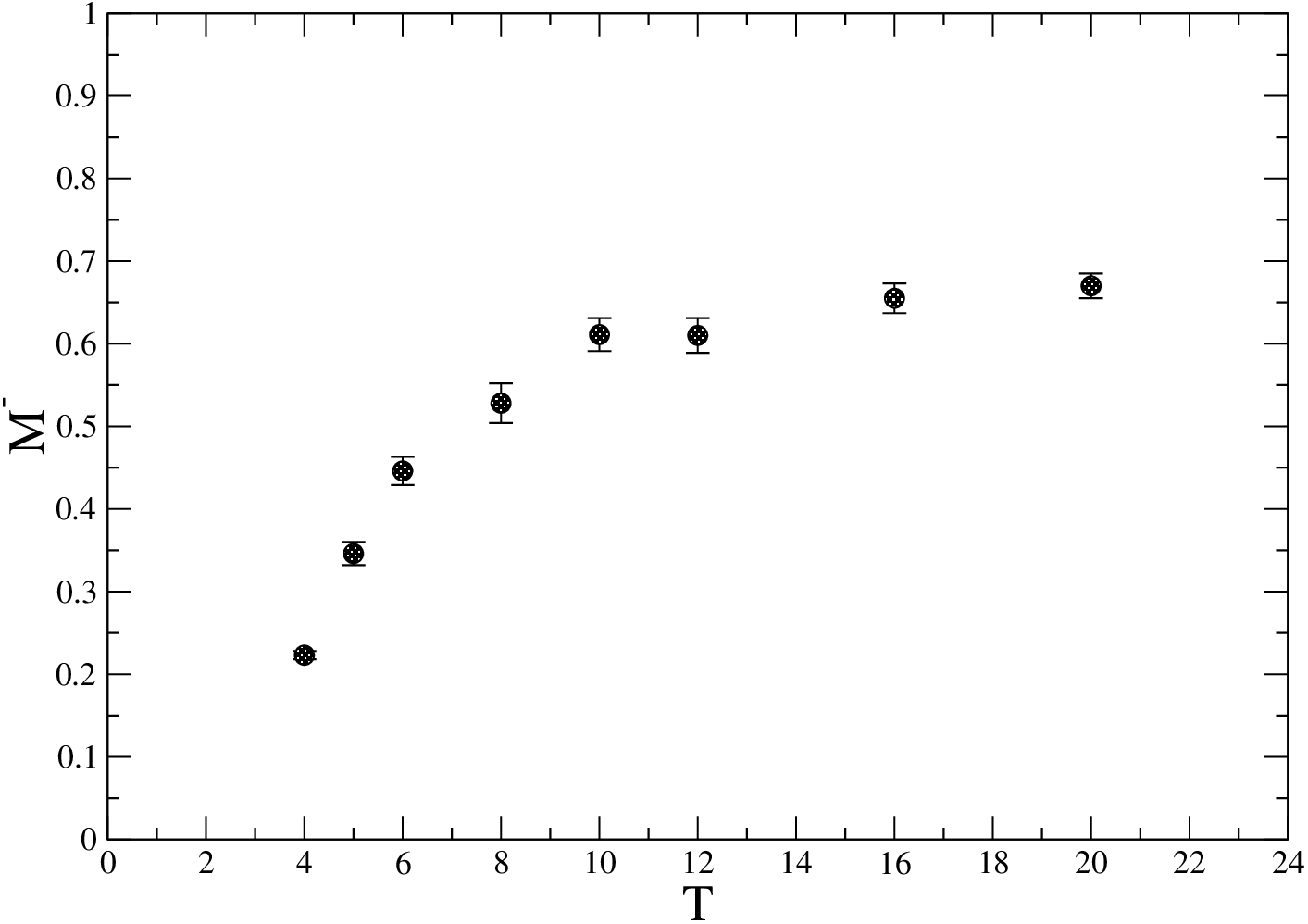}
\caption{The effective mass $M^-$ as a function of $T$.\label{fig:Mm}}
\end{center}
\end{figure}
\indent On each lattice the different determinations of $Z^s_{m,d}/Z$ are in good agreement, 
and the sum $(Z^+/Z+Z^-/Z)$ is always consistent with 1. For $Z^-/Z$  
a clear statistical signal is obtained for $m=2$ only, and the larger 
error at $m=1$ indicates that the exponential reduction of the noise 
is working as expected. To better appreciate the efficiency of the method, 
it is useful to define the quantity 
\be\displaystyle
M^{-} = -\frac{1}{T}\, \mathrm{Ln}\left[\frac{Z^-}{Z}\right]
\ee
whose values are reported in Table~\ref{tab:results}. With the exception
of the lattice ${\rm A}_4$, it is clear that $O(50)$ measurements are enough to 
obtain a precision on $M^{-}$ of the order of $5\%$ on both spatial volumes.
Sticking to the ${\rm A}$ lattices, the comparison of the relative errors 
on $M^{-}$ at $T=5,6,10,12$ and at $T=20$ indicates that the multi-level integration
indeed achieves an exponential reduction of the noise. 
The most precise determination of $Z^-/Z$  at each value of $T$ is plotted in 
Fig.~\ref{fig:ZmZp}. Its value decreases by more than five orders of magnitude 
over the time range spanned. The symmetry constrained 
Monte Carlo clearly allows to follow the exponential 
decay over many orders of magnitude, a fact which  represents one of the main results of 
the paper.\\
\indent The data in Table~\ref{tab:results} confirm the expectation that 
at these volumes the ratio $Z^-/Z$ suffers from large finite-size effects.
If we enforce the theoretical prejudice that
a single state with multiplicity 1 dominates $Z^-/Z$ for large $T$, then
$M^-$ can be interpreted as an effective parity-odd glueball mass, which should 
approach its asymptotic value from below. Indeed this is verified at both
values of $L$, as shown in Fig.~\ref{fig:Mm} for the A lattices. 

\section{Conclusions}
The exponential growth of the statistical error with the 
time separation of the sources is the main limiting factor 
for computing many correlators on the lattice by a standard 
Monte Carlo procedure. The integration scheme proposed here solves 
this problem by exploiting the symmetry properties of the underlying 
quantum theory, and it leads to an exponential reduction of 
the statistical error. In particular the cost of computing the energy of 
the lowest state in a given symmetry sector grows linearly with the 
time extent of the lattice.\\ 
\indent In extensive simulations of 
the SU(3) Yang--Mills theory, we have observed a definite exponential reduction
of the statistical error in the computation of the relative contribution 
of the parity-odd states to the partition function. The simulations needed at larger 
volumes and finer lattice spacings to provide a theoretically solid
evidence for the presence of a glueball state, and to precisely determine its mass 
are now feasible with the present generation of computers.\\
\indent Since the strategy is rather general, 
we expect it to be applicable to other symmetries and other field 
theories including those with fermions as fundamental degrees of freedom. 
In QCD, for instance, the very same problem occurs already in the computation
of rather simple quantities such as the energy of the vector meson resonance, 
and it becomes 
even more severe for the $\eta'$ and baryon masses. The approach presented
here offers a new perspective for tackling these problems on the lattice.\\
\indent The integration scheme described is yet another 
example of how the properties of the underlying quantum system,
namely the parity symmetry, can be exploited to design more efficient 
exact numerical algorithms for the computation of the dynamical 
properties of the theory.

\section*{Acknowledgments}
We thank Martin L\"uscher for many illuminating discussions, for a careful 
reading of the first version of the manuscript, and for the 
constant encouragement throughout the last year. Our simulations were 
performed on PC clusters at the University of Bern, at CILEA, and at 
the University of Rome ``La Sapienza''. We thankfully acknowledge the 
computer resources and technical support provided by all these institutions 
and their technical staff. 

\appendix 

\section{Numerical computation of $\;{\overline{\rm R}}$ \label{app:appA}}
In this Appendix we describe how the ratio 
${\overline{\rm R}}[V_{x_0+d},V_{x_0},r]$, defined in 
Eq.~(\ref{eq:rat2}), has been computed by a three-level algorithm. 
The partition function 
$\overline {\rm T}[V_{x_0+d},V_{x_0},r]$ is rewritten as 
\be
{\overline {\rm T}}\Big[V_{x_0+d},V_{x_0},r\Big] = \int 
 {\rm D}_4 [U]_\mathrm{sub} {\rm \bf D} [U_4]\; e^{-{\overline S[U,r]}}\; , 
\ee
where a second temporal link $U_4(y_0,\vec y)$ has been 
added to the standard degrees of freedom at each point of the time-slice 
$y_0=(x_0 +d-1)$. The subscript ``sub'' indicates the integration 
over the standard active-link variables of the thick time-slice 
$[x_0,x_0+d]$ with the spatial components $U_k(x)$ of the boundary fields 
fixed to $V_k(x_0,\vec {\bf x})$ and $V_k(x_0+d,\vec {\bf x})$ respectively. 
The modified action ${\overline S}[U,r]$ reads 
\be
{\overline S}[U,r] = S[U] + \frac{\beta}{6}\, (1-2r) \sum_{{\vec y}, k} 
{\rm Re}\Tr\Big\{U_{0 k}(y_0,\vec y) - U_{4 k}(y_0,\vec y) \Big\}\; ,
\ee
where $U_{0 k}(y)$ is defined in Eq.~(\ref{eq:placst}) and 
\be
U_{4 k}(y) = U_4(y_0,\vec y)\, U^\dagger_k(y_0+1,-\vec y - \vec k)\, 
U^\dagger_4(y_0,\vec y + \hat k)\,
             U^\dagger_k(y_0,\vec y)\; . 
\ee
If one defines the ``reweighting'' observable as 
\be
O[U,r+\varepsilon/2] = 
e^{{\overline S}[U,r+\varepsilon/2]-{\overline S}[U,r-\varepsilon/2]}\; ,
\ee
then the ratio ${\overline{\rm R}}[V_{x_0+d},V_{x_0},r]$ can 
be computed as its expectation value on the 
ensemble of gauge configurations generated with the action ${\overline S}[U,r+\varepsilon/2]$. In practice the average value of the observable $O$ is estimated by implementing 
the following three-level algorithm:
\begin{enumerate}
\item Generate a thermalized configuration with the action 
${\overline S}[U,r+\varepsilon/2]$ by spanning the sub-lattice 
with several sweeps of the update algorithm (see section 
\ref{sec:impl});
\item Compute an estimate of $\langle O \rangle$ by averaging over 
$n_0$ (level 0) configurations\footnote{Notice that when 
spatial links are kept fixed, the set of $U_0$ and $U_4$ factorize and 
are generated independently.} generated by keeping fixed 
all link variables with the exception of the links $U_0$ and 
$U_4$ on the time-slice $y_0$; 
      \item Repeat step 2 over $n_1$ (level 1) configurations generated by keeping 
      fixed all links of the sub-system with the exception of those on the 
      time-slice $y_0$, and average over the results obtained;
\item Repeat step 3 over $n_2$ (level 2) configurations generated by
      updating all links of the sub-lattice with the action 
      ${\overline S}[U,r+\varepsilon/2]$, and average over the results obtained.
\end{enumerate} 
At each level the numbers $n_0$, $n_1$ and $n_2$
of configurations generated are chosen to minimize the numerical cost 
required to reach the desired statistical precision. Their values depend on 
$d$ and $r$. In the simulations that we have carried out they range
in the intervals $n_0=12-50$, $n_1=50-120$ and $n_2=50-300$.

\bibliographystyle{h-elsevier}   
\bibliography{lattice}        
\end{document}